\documentclass[aip,preprint]{revtex4-1}
\usepackage{graphicx}
\usepackage{bm}
\usepackage{color}
\usepackage{hyperref}
\usepackage{ulem}
\usepackage{amsmath}
\usepackage{amssymb}
\usepackage{mathtools}
\usepackage{mathrsfs}
\usepackage{textcomp}

\usepackage{here}
\usepackage[dvipsnames]{xcolor}

\usepackage{lipsum}
\usepackage{multirow}
\usepackage{array,booktabs}
\newcolumntype{M}[1]{>{\centering\arraybackslash}m{#1}}

\begin{document}

\author{Upayan Baul}
\email{upayan.baul@physik.uni-freiburg.de}
\affiliation{Applied Theoretical Physics -- Computational Physics, Physikalisches Institut, Albert-Ludwigs-Universit\"{a}t Freiburg, Hermann-Herder Strasse 3, D-79104 Freiburg, Germany}
\author{Michael Bley}
\email{michael.bley@physik.uni-freiburg.de}
\affiliation{Applied Theoretical Physics -- Computational Physics, Physikalisches Institut, Albert-Ludwigs-Universit\"{a}t Freiburg, Hermann-Herder Strasse 3, D-79104 Freiburg, Germany}
\author{Joachim Dzubiella}
\affiliation{Applied Theoretical Physics -- Computational Physics, Physikalisches Institut, Albert-Ludwigs-Universit\"{a}t Freiburg, Hermann-Herder Strasse 3, D-79104 Freiburg, Germany}
\affiliation{Cluster of Excellence livMatS @ FIT - Freiburg Center for Interactive Materials and Bioinspired Technologies, Albert-Ludwigs-Universit\"at Freiburg, Georges-K{\"o}hler-Allee 105, D-79110 Freiburg, Germany}
\email{joachim.dzubiella@physik.uni-freiburg.de}

%
\title{Thermal compaction of disordered and elastin-like polypeptides: a temperature-dependent, sequence-specific coarse-grained simulation model}%

\begin{abstract}
Elastin-like polypeptides (ELPs) undergo a sharp solubility transition from low temperature solvated phases to coacervates at elevated temperatures, driven by the increased strength of hydrophobic interactions at higher temperatures. The transition temperature, or \textquoteleft cloud point\textquoteright, critically depends on sequence composition, sequence length, and concentration of the ELPs. In this work, we present a temperature-dependent, implicit solvent, sequence-specific coarse-grained (CG) simulation model that reproduces the transition temperatures as a function of sequence length and guest residue identity of various experimentally probed ELPs to appreciable accuracy. Our model builds upon the self-organized polymer model introduced recently for intrinsically disordered polypeptides (SOP-IDP), and introduces a semi-empirical functional form for the temperature-dependence of hydrophobic interactions. In addition to the fine performance for various ELPs, we demonstrate the ability of our model to capture the thermal compactions in dominantly hydrophobic intrinsically disordered polypeptides (IDPs), consistent with experimental scattering data. With the high computational efficiency afforded by the CG representation, we envisage that the model will be ideally suited for simulations of large-scale structures such as ELP networks and hydrogels, as well as agglomerates of IDPs. 
\end{abstract}
\maketitle
\section{Introduction}
Stimuli-responsive polymers are key ingredients in switchable \textit{smart} soft materials design -- materials whose physiochemical, and hence assembly properties can be critically altered by local environmental stimuli~\cite{Koetting2015,Gil2004,Liechty2010,Roth2017,Wei2017,Andreas}. A characteristic generic to responsive polymers is that they undergo a sharp and reversible conformational change over a relatively small window of the stimulus. Thermoresponsive polymers are a representative class of such stimuli-responsive polymers, and respond to temperature as the external stimulus. These polymers are diverse in composition, both biotic and abiotic, and have found wide-spread applications in the design of targeted switchable materials. Such applications include for example the design of surfaces with controllable wettability and stiffness, hydrogels, drug carriers and micellar nano-materials, controllers of gene expression and enzyme function, and tissue repair~\cite{Klouda2008,Pashkuleva_1,Mano2008,Ganta2008,Gibson2013,Alarcon2005,Khutoryanskiy2018book,Kim2017,Kyle_1}.

Elastin-like polypeptides (ELPs) are a special class of bio-inspired peptide-based thermoresponsive polymers, characterized by variants of the pentameric repeat motif (VPGXG). The guest residue X can in principle be any amino acid but proline~\cite{Roberts2015}. The most commonly studied or the canonical ELP (VPGVG)$_n$ derives from the evolutionarily conserved hydrophobic domains of tropoelastin, the precursor to extracellular matrix protein elastin~\cite{Vrhovski1998}. ELPs undergo a sharp solubility transition at lower critical solution temperature (LCST) from solvated phases at temperatures below to coacervates at temperatures above. The transition temperature, often referred to and measured as the cloud point $T_c$, depends most notably on sequence length (molecular weight), concentration of ELPs, and the identity of the guest residue X~\cite{Urry1997JPCB,Urry1991JACS,Urry1985Biopolymers,Urry1992Biopolymers,MacEwan2010,Meyer2002Biomacromol,Meyer2004Biomacromol,Tatsubo2018,Ribeiro2009,Jayaraman_SM,Jayaraman_ElpClp}. Strongly hydrophobic guest residues result in typically low $T_c$~\cite{Urry1992Biopolymers}. The same trend is observed upon increasing the sequence length, and concentration~\cite{Meyer2004Biomacromol}. 

The modulators of $T_c$ for ELPs are not limited to the physical attributes listed above, which are also the primary stimuli for synthetic responsive polymers such as PNIPAM~\cite{Tong1999,Furyk2006}. Several other external stimuli~\cite{Urry1997JPCB,Chilkoti2006CurrOpin}, such as ion concentration and specificity~\cite{Cremer2008JPCB}, pH~\cite{Ribeiro2009}, and the presence of denaturants, co-solvents and co-solutes,~\cite{Cremer2006Biomacromol,ELP_Cononsolvency,Urry1991JPC} etc., can appreciably change the $T_c$ for ELP sequences, imparting them with substantially higher, and also more tunable responsiveness to changes in local environment~\cite{MacEwan2010}. It should be noted that polymers such as PNIPAM and corresponding synthetic co-polymers can also be responsive to the above stimuli, albeit to a lesser degree of tunability~\cite{Cremer2007JPCC,Cremer2005JACS,Karg2008,Zhang2007}. This heightened receptivity together with additional advantages such as bio-compatibility, precise control over molecular weight in production~\cite{Meyer2002Biomacromol,Kurihara2005,McMillan99}, recombinant peptide synthesis~\cite{Lecommandoux2}, tunable co-polymerization~\cite{Conti02,Chilkoti02diblock,Bitton17,Lecommandoux1,Kiick}, and complex self assembled structures~\cite{MacEwan2010,MacEwan2017,Champion_1,Nose_1}, makes ELPs ideally suited for applications such as protein sorting, tissue repair, and drug carrier design~\cite{Lecommandoux1,MacEwan2010,Le2019,Desp16}.

In addition to their potential for industrial and therapeutic applications, ELPs, owing to their low sequence complexity, have historically been used as templates for studying the temperature-dependence of inter-amino acid interactions. In notable early studies, Urry \textit{et al.} systematically investigated the influence of the guest residue identity on $T_c$, hence the overall hydrophobicity of ELP sequences~\cite{Urry1985Biopolymers,Urry1991JACS,Urry1992Biopolymers}. The different amino acids were accordingly ranked based upon their observed hydrophobicities~\cite{Urry1992Biopolymers}. Subsequent experimental studies spanning over two decades have elucidated the influence of diverse stimuli on the LCST behavior of diverse ELP sequences~\cite{Urry1997JPCB,Chilkoti2006CurrOpin,Cremer2008JPCB,Ribeiro2009,Cremer2006Biomacromol,Urry1991JPC}. A mechanistic understanding of the LCST in ELPs, however, is difficult to ascertain from experimental measurements alone. At the heart of the problem lies the inability of conventional experimental techniques to probe the conformational ensembles of disordered polypeptides at high spatial resolution~\cite{Delaforge2018,Deniz2010}. In experimental studies $T_c$ is conventionally obtained using macroscopic measurements of turbidity profiles through the \textquoteleft cloud-point\textquoteright~of ELP solutions~\cite{CldPt1,Bitton17}. 

Computer simulations using all-atom resolution and state-of-the-art force fields have been instrumental in studying the room-temperature, as well as temperature-dependent structure, and mechanism of LCST transition in ELPs~\cite{Rauscher17,Ying18,Zhao2016,Ying14,Marx04,Glaves08, Kremer}. For example, simulations have revealed that contrary to Urry \textit{et al}'s $\beta$-spiral model~\cite{Urry1981JACS}, the high temperature phases of ELPs are substantially disordered, and resemble the molten \textquoteleft collapsed globule\textquoteright~states observed in proteins~\cite{Rauscher17}. Insights have been gained on the significance of desolvation-mediated-attraction among hydrophobic amino acid side chains~\cite{Ying14}, transient hydrogen bonding interactions along the polypeptide backbone~\cite{Zhao2016,Marx04}, and the disorder-promoting influence of proline residues~\cite{Glaves08}, etc. Owing to the inherent computational cost associated with explicit-solvent simulations, the length (and time) scales that can be studied by all-atom simulations is limited. These simulations accordingly cover smaller sequence lengths than studied experimentally. In order to study larger length scale structures such as ELP networks and hydrogels, it is rewarding to use more coarse-grained (CG) simulation models~\cite{Kremer,MittalTdep}, while also retaining the sequence-specific and temperature-dependent properties pertaining to the different ELP sequences. The development of such an {\it explicitly temperature-dependent}, i.e., temperature-transferable CG model~\cite{MPlathe,Voth,Nico,MittalTdep} is the primary objective of the present study.

Temperature-dependence of the dimension and phase behavior of natively disordered polypeptides is not specific to ELPs alone. It depends strongly on sequence composition and directionality, as demonstrated in recent years by studies on phase demixing in biological systems~\cite{RevMittag2018BioChem,TaylorCell2015,ParkerMolCell2015,MittagJACS2016,BrangwynneScience09,RosenNature2012,CruzPsep,SheaPsep}. Predominantly polar sequences remain expanded, or under good-solvent conditions irrespective of the solution temperature. A high content of charged and aromatic amino acids is observed to promote coacervation below upper critical solution temperatures (UCST), signifying the role of multivalent electrostatic and $\pi$-$\pi$ or cation-$\pi$ interactions in the UCST behavior~\cite{RevMittag2018BioChem,Quiroz2015,ParkerMolCell2015,TaylorCell2015,FawziMolCell2018}. LCST behavior, or compaction at high temperatures can be mapped to the abundance of hydrophobic (and hydrophobic aromatic) residues, but with low charge content in the sequences~\cite{RevMittag2018BioChem,Quiroz2015}. It is, however, intriguing to observe that even after accounting for the (expected) temperature-dependences of the above mentioned interactions using theory and detailed molecular simulations, accounting for the temperature-dependent compaction of IDP sequences to quantitative agreements with experimental observations still remains elusive~\cite{AbsinthTdep}. This conundrum was highlighted by Skep\"{o} \textit{et al.}, phrased through the highly pertinent question,-- \textquotedblleft \textit{Is it possible for the currently available simulation methods to accurately mimic the experimental temperature induced structural changes in IDPs?}\textquotedblright~\cite{Skepo2019JCTC}. Recent work by Mittal \textit{et al.} provides a viable strategy to a temperature - dependent parameterization of a CG model of IDPs, where the authors suggest refinements of the strength of pairwise Lennard-Jones interactions among all amino acids using parabolic functions in temperature~\cite{MittalTdep}. The empirical parabolic form is motivated by statistical mining of protein structures by Abeln \textit{et al}~\cite{vanDijk2015}.

With the focus of the current study being LCST transitions in long ELP chains, the temperature-dependence in our model is dedicated primarily to interactions among hydrophobic residues. To this end, we propose and systematically parameterize in this report a simple, semi-empirical functional form for the strength of pairwise interactions among hydrophobic moieties as a function of temperature. We demonstrate that our functional form results in quantitative reproduction of $T_c$ in hydrophobic ELPs for the first time in CG simulations, and also reflects the high temperature compactions of more complex IDP sequences. Explicit parameters are reported for use with the self-organized polymer model for intrinsically disordered polypeptides (SOP-IDP)~\cite{SOPidp,SOPidpABeta}, but the generic applicability of the functional form to other CG models holds by construction. When applicable, we also consider the temperature-dependence of the dielectric constant of water, and inverse Debye screening length of salt solutions to account for the temperature-dependence of electrostatic interactions to a first approximation~\cite{TdepDielctricMaryott1956}.

The rest of the manuscript is organized as follows. The methods section starts with a brief description of the temperature-independent SOP-IDP model~\cite{SOPidp}, followed by a detailed description of the incorporation of temperature-dependence into the energy function. All computational observables used in the manuscript are also outlined, together with simulation protocols. In the results section, we first describe the parameterization of the model using only two sequence lengths of the canonical ELP sequence (VPGVG)$_n$. The model is then applied to a detailed study of the temperature induced structural changes in diverse uncharged ELP sequences. In the final section of the results, we test our simulation model against reported experimental results for two IDPs well known for their complex transient structure forming propensities. 
\section{Methods}
In order to incorporate temperature-dependence of inter-residue interactions in a CG model, it is practical and convenient to develop a model that describes the ambient-temperature phases of disordered polypeptides for diverse sequence compositions. The recently developed SOP-IDP model has been shown to reproduce the radii of gyration ($R_g$), as well as the small angle X-ray scattering (SAXS) profiles at room temperatures for diverse sequence compositions, including highly charged, fully polar, and predominantly hydrophobic sequences~\cite{SOPidp}. We thus chose the SOP-IDP model, devoid of any explicit temperature-dependence of inter-residue interactions, as the template to build the temperature-dependence upon. 

\begin{figure}[!htb]
\includegraphics[width=0.44\textwidth]{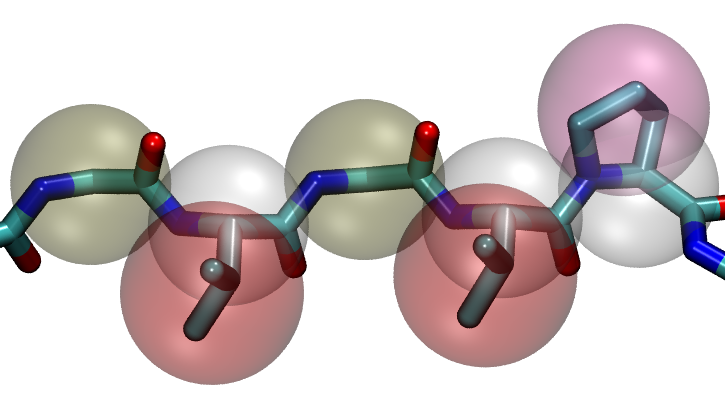}
\vspace{-0.5cm}
\caption{{\footnotesize Schematic representation of the mapping of a peptide segment, --GVGVP-- (from left to right, H-atoms not shown for clarity) onto the SOP-IDP model. Backbone beads for valine (V) and proline (P) residues are shown using transparent silver spheres, while the corresponding side-chain beads are shown using red and purple transparent spheres respectively. The pale brown transparent spheres along the peptide backbone represent glycine (G) residues described by a single bead.}}
\label{fgr:CG-scheme}
\end{figure}

In the current implementation of the SOP-IDP model an amino acid residue is represented by two beads, a backbone bead and a side-chain bead. Glycine, with its side chain comprising of only a single H-atom, is the only exception to this rule, and is described using a single bead. Figure~\ref{fgr:CG-scheme} shows a schematic representation of the mapping of a peptide segment onto the CG SOP-IDP model. 

The energy function for the SOP-IDP model is given by the equation\\ 
\begin{eqnarray}
  E & = &- \sum_{i=1}^{N_B} \frac{k}{2} R_0^2 \log \left(1 - \frac{(r_i - \sigma_i)^2}{R_0^2} \right) + \sum_{i=1}^{N_{loc}} \epsilon_{loc} \left(\frac{\sigma_i}{r_i} \right) ^6 + \sideset{}{'} \sum_{i,j}^{} \frac{e_i e_j \exp \left( -\kappa r_{ij} \right) }{\varepsilon r_{ij}} \nonumber \\
  & +& \sum_{i=1}^{N_{BB}} \epsilon_{BB} \left[ \left( \frac{\sigma_i}{r_i} \right)^{12} - 2\left( \frac{\sigma_i}{r_i} \right)^{6} \right] + \sum_{i=1}^{N_{BS}} \epsilon_{BS} \left[ \left( \frac{\sigma_i}{r_i} \right)^{12} - 2\left( \frac{\sigma_i}{r_i} \right)^{6} \right] \nonumber \\
  & +& \sum_{i=1}^{N_{SS}} \epsilon_{SS} \left[ \left( \frac{\sigma_i}{r_i} \right)^{12} - 2\left( \frac{\sigma_i}{r_i} \right)^{6} \right]. \label{eqn:Ham}
\end{eqnarray}
%
The first term represents bonded interactions described by finite non-linear elastic (FENE) potential, where $N_B$ is the total number of bonds in the system. The second, purely repulsive term is only active between bead pairs that are not covalently bonded, but belong to residues separated by $\leq$ 2 along the sequence. The total number of such bead pairs is denoted by $N_{loc}$. The third term is a screened Coulomb potential accounting for electrostatic interactions among charged beads. The parameters $\kappa$ and $\varepsilon$ in this term stand for the inverse Debye length, and the dielectric constant, respectively. The charge of an amino acid is assigned to the side-chain bead. Details on the significance these terms, in the context of both folded proteins and disordered sequences, can be found elsewhere~\cite{SOPidp,SOPidpABeta,SOP2012Liu,SOP2015Reddy}.

%
The final three terms in eq~\ref{eqn:Ham} account for the inter-bead Lennard-Jones (LJ) interactions in the model, and incorporate sequence-specificity beyond electrostatic interactions. These interactions are only active among bead pairs that do not interact through the first two terms in eq~\ref{eqn:Ham}. In order of appearance, they represent the pairwise interactions among backbone - backbone, backbone - side-chain, and side-chain - side-chain beads (total $N_{BB}$, $N_{BS}$, and $N_{SS}$ such pairs respectively). The van der Waals (vdW) radius for a bead pair ($\sigma_i$) is the sum of the vdW radii of the interacting beads. To avoid any ambiguity with notations, it is worth noting that $\sigma_i$ in our notation represents the separation corresponding to the minimum of the pairwise LJ interaction, often denoted by $r_i^{min}$ in conventional notation for LJ interactions ($\sigma_i^{\mathrm{SOP-IDP}} = r_i^{min} = 2^{1/6} \sigma_i^{LJ}$). 
The term $\epsilon_{SS}$ has the further explicit form $\epsilon_{SS} = \epsilon_{SS}^0 \left|\epsilon_i - 0.7\right|$, where the amino acid pair specific parameter $\epsilon_i$ is obtained from the knowledge-based Betancourt-Thirumalai statistical potential, and is used to reweight the inter-residue interactions at ambient temperature ($T \approx 298$ K)~\cite{BetTh}. The energy scales of these interactions are set by parameters $\epsilon_{BB}$, $\epsilon_{BS}$, and $\epsilon_{SS}^0$, parameterized to describe disordered polypeptide sequences at ambient temperatures~\cite{SOPidp}. Description of the individual bead properties and numerical values of parameters in eq~\ref{eqn:Ham} can be found in Tables S1-S3 in the \textit{Supplementary Information} (SI).

\subsection{Temperature-dependence of hydrophobic inter-residue interactions}
Experiments, theory, and simulations have shown that the solvation of hydrophobes in water has a strong dependence on both temperature and size~\cite{Kauzmann1959,Chandler2000PNAS,Chandler2002JPCB,Garde_1,Garde_2,Garde_3}. For isolated hydrophobic moieties having dimensions similar to amino acids, within the temperature range of 273 K to 400 K the hydrophobicities first increase with temperature, and then decrease. This temperature-dependence has origins in the changes in structural dynamics of the solvating water molecules, and the entropy of solvation changes sign over the temperature interval~\cite{Chandler2000PNAS,Murphy1994,Garde_2}. Similar behavior has recently been evidenced from statistical mining of protein structures at different temperatures~\cite{vanDijk2015}. However, while the temperature-dependence of hydrophobic interactions is widely acknowledged to be of critical importance in the phase behavior of both folded and disordered polypeptide sequences, and generally for thermoresponsive hydrophobic polymers~\cite{Schellman1997,Reinhardt2013,ElcockAA}, explicit temperature-dependence of pairwise hydrophobic interactions is not conventionally used in CG models~\cite{Schulz2015,TozziniCurrOpin2005,TozziniAccChemRes2010,TozziniQRevBioPhys2010,MARTINIcg,ElcockMARTINI} because of the challenges in the consistent thermodynamic mapping of the atomistic degrees of freedom to the CG representation~\cite{MPlathe,Voth,Nico}.

For our knowledge-based SOP-IDP Hamiltonian (eq 1), we now construct a semi-empirical functional form for incorporating the temperature-dependence into the LJ potential as used here. Starting point is the observation that the hydrophobic attraction increases with temperature and thus the dimensionless second virial coefficient $B_2(T)/\sigma^3$ of hydrophobic pair potentials $U(r,T)$ (note the explicit dependence on $T$) between two monomers is negative and a decreasing function of temperature~\cite{Ludemann,Chaudhari20557}. Indeed for a thermosensitive polymer, it was observed that $B_2$ is simply linear in $T$ over a sufficiently large temperature interval (ca.~30 K) covering the transition temperature ($T_c$)~\cite{Reinhardt2013}. The increase of the hydrophobic pair (monomer-monomer) attraction with $T$ was also observed in explicit-water simulations of the thermosensitive polymer polyethylene-glycol (PEG), for which also an explicitly temperature-dependent CG model was devised recently~\cite{Chudoba2017PEG}. Hence, we assume that over a small temperature window it can be approximated by an expansion in powers of $T$ around some reference temperature, where we truncate the expansion after second order 
\begin{equation}
 B_2(T)/\sigma^3 = a_0 - a_1 T - a_2 T^2 \,.
 \label{eqn:B2Tdep}
\end{equation}
The coefficient $a_1$ we assume is a universal constant for water as solvent and not dependent on the type of hydrophobic monomer~\cite{Reinhardt2013}. For $U(r,T)$, being parameterized as a LJ pair potential, we seek now to obtain a functional form for the temperature-dependence of the pairwise interactions among hydrophobic amino acids. Analytical treatment of $B_2(T)$ for the LJ potential, however, poses its own challenges~\cite{Vargas2001}. For simplicity, we use the Mayer function definition 
\begin{equation}
 \frac{B_2(T)}{\sigma^3} = -\frac{2 \pi}{\sigma^3} \int_0^{\infty} \left[e^{-V(r)/k_BT} - 1\right]r^2dr
 \label{eqn:B2}
\end{equation}
and solve it using a shifted square well potential $V(r,T)$ with a temperature-dependent well depth $\epsilon(T)$, thus arriving at a simple, functional form for the LJ attraction energy $\epsilon(T)$ between the CG beads, given by (see SI for derivation)
\begin{equation}
 \epsilon(T) = k_BT \ln \left[1 + c (T - T_0) + d (T - T_0)^2 \right].
 \label{eqn:EpsT}
\end{equation}
This differs from the parabolic forms suggested in previous studies using empirical means~\cite{vanDijk2015,MittalTdep}. Through physical justifications alone, constraints apply to the possible values of $c$ and $d$, pertaining to pairwise hydrophobic bead interactions. First, $|c| \gg |d|$, $c > 0$ and $d < 0$ to account for increase in hydrophobicity with $T$, followed by decrease. Moreover, the envisaged universality of $a_1$ in eq~\ref{eqn:B2Tdep} dictates that $c$ should also be universal, with $T_0$ as a free parameter specific to a pair of hydrophobes. However, considering also the length-scale dependence of hydrophobic interactions, it is more reasonable to expect that $c$ should be universal to the extent of only small differences in the dimensions of hydrophobic species.

To incorporate temperature-dependence in our Hamiltonian (eq~\ref{eqn:Ham}), we use the functional form of eq~\ref{eqn:EpsT} to obtain $\epsilon_{SS}(T)$ for pairwise interactions among hydrophobic side chains. Specifically, temperature-dependence is considered in pairwise interactions among the side-chain beads corresponding to residues valine (Val, V), proline (Pro, P), leucine (Leu, L), isoleucine (Ile, I), methionine (Met, M), alanine (Ala, A), phenylalanine (Phe, F), tyrosine (Tyr, Y), tryptophan (Trp, W), and neutral histidine (His, H). With the original SOP-IDP model parameterized at $T=298$ K, the $T_0$ specific to pairwise interactions, $T_0^{ij}$, are estimated using the linear relation 
\begin{eqnarray}
\epsilon_{SS}^{ij}(T=298\,\mathrm{ K}) = \epsilon_{SS}^{ij}(\mathrm{SOP-IDP}) \nonumber\\
 \sim  k_B (298\,\mathrm{K}) \ln \left[1 + c (T - T_0^{ij}) \right] 
 \label{eqn:T0est}
\end{eqnarray}
where $\epsilon_{SS}^{ij}(\mathrm{SOP-IDP})$ is the corresponding interactions strength between the pair in the temperature-independent SOP-IDP model~\cite{SOPidp}. Numerical values for pairwise $T_0^{ij}$ are listed in Table S4 in the SI. 

For the electrostatic interaction term in eq~\ref{eqn:Ham}, the dielectric constant ($\varepsilon$) and the Debye screening length ($\kappa^{-1}$) are considered temperature-dependent. This term only contributes in simulations of IDPs in the manuscript, where charged amino acids are present along the sequence. Specifically, for a monovalent salt solution, $\kappa^{-1} = \left(\frac{\varepsilon \varepsilon_0 k_BT}{2\times 10^3 e^2 N_A C}\right)^{1/2}$, where $\varepsilon_0$ is the permittivity of free space, $e$ is electronic charge, $N_A$ is Avogadro number, and $C$ is molar concentration of salt. Specific values for simulations at given temperatures were calculated using an online tool~\cite{InvDebye}, using temperature-dependent~\cite{TdepDielctricMaryott1956} $\varepsilon$ values, and assuming a physiological monovalent salt concentration of 0.15 M. The $\kappa^{-1}$ values at select simulation temperatures are noted in Table S3 in the SI, and ranges between 7.89 \AA~and 7.54 \AA~in the temperature interval 288 K to 353 K. 
\subsection{Definitions of observables}
\textbf{Hydrodynamic radius ($R_h$) and radius of gyration ($R_g$):} The $R_h$ of a polymer chain is defined as the radius of an effective hard sphere, that has the same effective center of mass diffusivity as the polymer. From the simulated trajectories, we calculated $R_h$ using the conventionally used Kirkwood double-sum formula~\cite{doiBook},
\begin{equation}
 \frac{1}{R_h} \,=\, \frac{1}{N^2} \left \langle \sum_{i \neq j} \frac{1}{r_{ij}} \right \rangle 
 \label{eqn:Rh}
\end{equation}
known to reproduce the experimental Stoke's radius (as measured, for example, using DLS experiments~\cite{Stetefeld2016}) to within 10\% accuracy~\cite{Mansfield2007}. Additionally, in the SI, we also report $R_h$ calculated using a relation proposed recently by Nygaard and co-workers~\cite{NygaardRh} for estimating the $R_h$ of disordered polypeptides from $R_g$ and sequence length. This we refer to as $R_h^{\mathrm{Nyg}}$, and the calculation of it is described in the SI.

The $R_g$ values were calculated using~\cite{doiBook}
\begin{equation}
 R_g \,=\, \sqrt{ \frac {1}{N}\Bigl\langle\sum_{i=1}^{N}(\vec{r}_{i}-\vec{r}_{CM})^{2}}\Bigr\rangle \,\,\,.
 \label{eqn:Rg}
\end{equation}
In equations~\ref{eqn:Rh} and \ref{eqn:Rg}, $N$ represents the total number of beads in the polymer. The pairwise distances are denoted by $r_{ij}$, and position vectors for individual beads and the center of mass are denoted by $\vec{r}_{i}$ and $\vec{r}_{CM}$, respectively.\\

\noindent
\textbf{Definition of $T_c$:} 

The cloud point temperature $T_c$ is typically used as a measure of the CST of a polymer solution~\cite{Urry1997JPCB,Meyer2002Biomacromol}. Note that a first-order CST is only defined for polymer solutions in the thermodynamic limit~\cite{Ivanov1998,Schnabel2011}. The quantification of the transition of a large, many chain system is out of our computational reach, even using a coarse-grained model. We thus use the fact that the coil-to-globule transition temperature of a single polymer chain is a good marker for the LCST of polymer solutions, which has been experimentally justified, for example, for the thermoresponsive PNIPAM polymer~\cite{chiwu}. The transition from low-temperature expanded coil-like states to the high-temperature collapsed globular states is reflected in the probability distributions of $R_g$, denoted by $P(R_g)$ (Figure~\ref{fgr:ELPV_param}). The coil-to-globule transition for simple and highly flexible single polymer chains is known to be continuous, i.e., no explicitly bimodal two-state behavior can be observed for it. Two-state behavior can only be observed in polymer models by accounting for sufficient internal degrees of freedom~\cite{Maffi2012}, or by increasing stiffness~\cite{grosberg}. Using the calculated distributions for a polymer with $N_b$ bonds forming the backbone chain at simulated temperature $T$, $P(R_g;T)$, it is possible to estimate the probabilities of globular- ($P_g$), and coil-like ($P_c$) states at the given temperature using
\begin{eqnarray}
 P_g(T) &=& \int_{R_g \leq R_g^{ic}}  P(R_g;T) dR_g \nonumber \\
 P_c(T) &=& \int_{R_g > R_g^{ic}}  P(R_g;T) dR_g \,.
 \label{eqn:Tc}
\end{eqnarray}
Here $R_g^{ic}$ is a threshold value separating collapsed (globule-like) and swollen (coil-like) states, which we choose to be the theoretical radius of gyration of an equivalent ideal chain (or Gaussian chain), $R_g^{ic}\,=\, \sqrt{N_b}b/\sqrt{6}$. The choice ensures that we have the correct limiting behavior of the transition for infinitely long chains, where the $\theta$-point coincides with the the critical point for the phase separation of the polymer solution~\cite{wang2014,degennesbook,rubinsteinbook}. We calculated from our simulations that the average bond length, $b$, has only a weak dependence on $T$ ($\Delta b \sim 10^{-3}$ nm), and is well described by $b=0.41$~nm. The transition temperature $T_c$ is then naturally defined by the equality of the two states $P_g(T\,=\,T_c) \,\approx\, P_c(T\,=\,T_c)$. Similar definitions have also been used in the experimental studies of thermal denaturation of folded proteins~\cite{Schuler2013JACS}. \\

\noindent
\textbf{Polymer ensemble shape parameters:} The effective shape of a polymer conformation, as well as the distribution of shapes in a polymer conformational ensemble, can be characterized using distributions of variables $\Delta$ and $S$, defined as~\cite{ShapeRNA,Dima04JPCB} 
\begin{subequations}
\begin{align}
 \Delta \, &= \, \frac{3}{2} \frac{\sum_{i=1}^3 \left(\lambda_i - \bar{\lambda}\right)^2}{\left(3\bar{\lambda}\right)^2}\label{eqn:shapeD} \\
 S \, &= \, \frac{\prod_{i=1}^3 \left(\lambda_i - \bar{\lambda}\right)}{\bar{\lambda}^3}\,\,,
 \label{eqn:shapeS}
 \end{align}
\end{subequations}
where $\lambda_i$ are the eigenvalues of the gyration tensor, and $\bar{\lambda} = \left(\lambda_1 + \lambda_2 + \lambda_3 \right) / 3$~\cite{ShapeRNA}. The parameter $\Delta$ is a measure for the asphericity of conformations, and takes values between 0 (spherical) and 1 (linear). The parameter $S$ is negative for oblate ellipsoids and positive for prolate ellipsoids. The allowed values of $S$ follow the bound $-\frac{1}{4} \leq S \leq 2$~\cite{polymer_shape}.\\

\noindent
\textbf{Simulated SAXS profiles:} SAXS profiles for the scattering intensity $I_q$ for wave-vector $q$ from the simulated trajectories were computed using the Debye formula~\cite{PutnamRevSAXS}
\begin{equation}
 I_{q}\,=\,\Bigl\langle \sum_{i=1}^{N} \sum_{j=1}^{N}  f_i(q) f_j(q) \frac{\sin(q r_{ij})}{q r_{ij}} \Bigl\rangle,
 \label{eqn:Debye}
\end{equation}
where $N$ is the total number of beads in the polymer and $r_{ij}$ are pairwise distances. Optimized $q$-dependent form factors $f(q)$ derived using electron-density matching for two bead per residue CG representations of amino acids are available in literature~\cite{QdepFF}. The same (Table S1 in the reference~\cite{QdepFF}) were used in our analysis. For dilute solution, the $I(q)$ calculated using eq~\ref{eqn:Debye} is identical to the single chain form factor.
\subsection{Simulations and data analyses}
Underdamped Langevin dynamics simulations in the canonical (NVT) ensemble were carried out using the LAMMPS molecular dynamics simulator~\cite{lmp1,lammpsLangevin}. All simulations were carried out at vanishingly small concentrations by assigning very large box dimensions, such that the simulated molecule can not interact with its periodic images. Simulation time-step was chosen to be 30 fs. Unless specified otherwise, each simulation trajectory was equilibrated for $10^7$ timesteps, followed by production simulations for $10^8$ timesteps during which $10^4$ conformations were stored at equal intervals for analysis. For all systems a minimum of six independent simulations were performed. The software Visual Molecular Dynamics (VMD) was used for visualization and rendering of representative conformations~\cite{vmd}. Truncation and shift was used for all non-bonded interactions described in eq~\ref{eqn:Ham}. The cut-off distance for LJ interactions was 2.4 nm. The screened Coulomb interaction cut-off was chosen to be $4\kappa^{-1}$.

In the manuscript, quantitative comparisons of simulated and experimental data have been attempted with four experimental reports. In the following, we briefly outline the key experimental conditions such as pH, ionic strength and buffer conditions from the respective reports when available. The simulated $T_c$ for (VPGVG)$_n$ is compared to turbidity profile measurements by Meyer and Chilkoti~\cite{Meyer2004Biomacromol}. The publication does not report any of the above conditions. Our simulations for these chains, which contain no charged amino acids, are also unaffected by salt concentration and pH through construction. For IDP p53-IDR, the SAXS profile at 293 K is compared with experiments by Fersht \textit{et al.}~\cite{p53saxs} (25 mM phosphate, pH 7.2, 0.15 M NaCl, 5 mM DTT, 5\% (vol/vol) glycerol), and $R_h$ is compared with dynamic light scattering (DLS) measurements by Whitten \textit{et al.}~\cite{p53DLS}. The latter does not report DLS experiment conditions explicitly, whence physiological conditions are assumed. For IDP hTau40, $R_g$ and $I(q)$ profiles are compared with SAXS measurements by Tenenbaum \textit{et al}~\cite{saxsTauTdep,saxsTauTdep2012}. The authors reported that protein powder was reconstituted in 50 mM MES, pH 6.8, 0.1 M NaCl and 0.5 mM EGTA, concentrated at 2 mg/ml in 0.1 $\times$ phosphate buffered saline solution of ionic strength 0.02 M at pH 7.4 (among other conditions), and finally centrifuged and filtered~\cite{saxsTauTdep}. Owing to the simplicity of our energy function and simulation protocol, experimental conditions such as buffer and denaturants can not be accounted for in the simulations. Consistent with all reported pH values, histidine was considered to be neutral in simulations. Charges for other amino acids are reported in the SI. All simulations were performed considering 0.15 M monovalent salt concentrations to evaluate Debye screening lengths $\kappa^{-1}$. In our simple protocol of using screened electrostatic interactions, the accounting for charged bead interactions is only approximate.

\section{Results and discussion}
\subsection{Assignment of universal parameters in the model: LCST in (VPGVG)$_n$}
%
%
\begin{figure*}[h]
\includegraphics[width=1.0\textwidth]{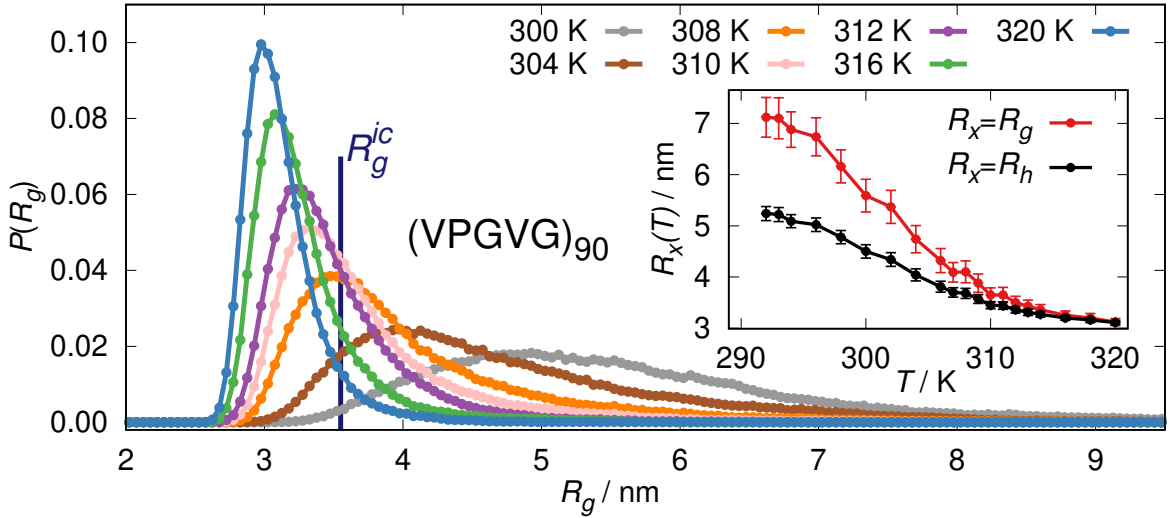}
\caption{{\footnotesize Probability distributions of radius of gyration ($R_g$) for the ELP (VPGVG)$_{90}$ at different temperatures across the observed $T_c$ ($\approx$ 310K). The broad distributions at low temperatures can clearly be observed to systematically shift towards sharper distributions about smaller mean $R_g$ values, as the simulation temperature is increased. The vertical line shows the $R_g$ for an equivalent \textit{ideal chain} polymer ($R_g^{ic}$, see Methods). The inset plot shows the temperature-dependences of mean $R_g$ and $R_h$ for the ELP.}} 
\label{fgr:ELPV_param}
\end{figure*}

In notable experiments, Meyer and Chilkoti studied the LCST transitions for the canonical ELP sequence (VPGVG)$_n$ (along with other sequences of lower hydrophobicities) for various sequence lengths ($N = 5n$)~\cite{Meyer2004Biomacromol}. The observed $T_c$ values at the very low experimental solution concentrations are ideally suited for parameterization of our model, allowing direct comparisons of simulated and experimental results, as also done previously~\cite{Zhao2016}. We used the $T_c$ values at the lowest reported ELP concentration of 1 $\mu$M from the aforementioned study (see supplementary figure B therein) for (VPGVG)$_{60}$ and (VPGVG)$_{90}$ to determine the universal parameters $c$ and $d$ (eq~\ref{eqn:EpsT}) in our model.

\begin{figure*}[!htb]
\includegraphics[width=1.0\textwidth]{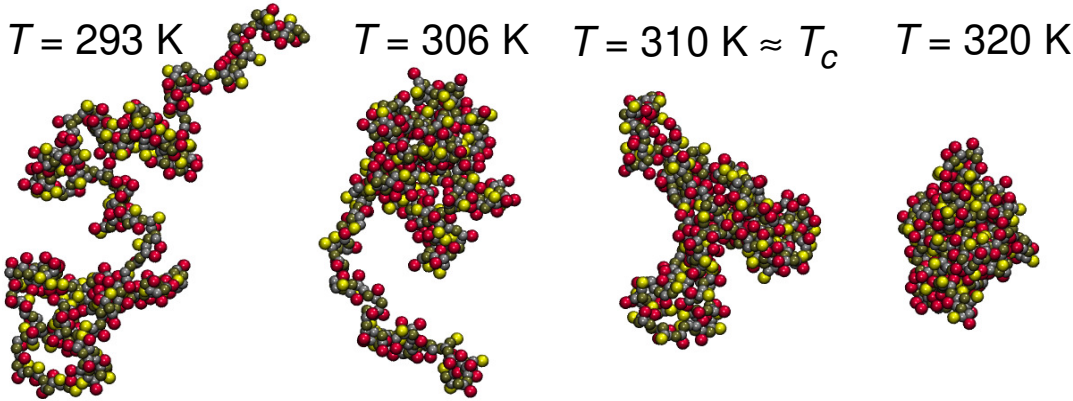}
\caption{{\footnotesize Representative snapshots showing the progressive collapse of a (VPGVG)$_{90}$ sequence as simulation temperature ($T$) is increased from below to above LCST ($T_c \approx 310 K$). The relative dimensions are true to scale.}}
\label{fgr:Snapshots}
\end{figure*}
Initial estimates of reasonable ranges for $c$ and $d$ were aided by (i) the constraints on $c$ and $d$ discussed in the Methods section, (ii) novel insights from simulations~\cite{Chan2004JACS}, and theoretical results on length-scale dependence of temperature-dependent hydrophobicity~\cite{Chandler2000PNAS,Chandler2002JPCB}. Details are included in the SI. Note that for justified choices of $c$ and $d$, progressive compaction of a hydrophobic ELP sequence at elevated temperatures between $\sim$(290 K - 330 K) is a naturally emergent property of the functional form of $\epsilon_{SS}(T)$ (cf. eq~\ref{eqn:EpsT}). This is highlighted in the normalized probability distributions of $R_g$, $P(R_g)$, shown at different values of simulation temperature in Figure~\ref{fgr:ELPV_param}. A visual examination of the plots shown for (VPGVG)$_{90}$ already makes it apparent that at low temperatures of $\sim$ 300 K the conformational ensemble results in a broad distribution of $R_g$ values with almost all conformations contributing to $R_g$ values greater than the ideal chain $R_g$, $R_g^{ic}$. As temperature is increased, the distribution as well as the mean value (shown in inset of Figure~\ref{fgr:ELPV_param}) shifts to smaller values of $R_g$, ultimately resulting in the reverse scenario where almost all conformations are compact, having $R_g < R_g^{ic}$. The progressive compaction is highlighted in Figure~\ref{fgr:Snapshots} using representative snapshots of polymer conformations. We used such simulated distributions, together with the definitions of $T_c$ from eq~\ref{eqn:Tc} to estimate the LCST from our simulations for different values of $c$ and $d$. Simultaneously comparing the $T_c$ from simulations with the experimental cloud points ($T_c^{exp}$) for (VPGVG)$_{60}$ and (VPGVG)$_{90}$ ($T_c^{exp}\,=\,$ 317 K and 310 K respectively~\cite{Meyer2004Biomacromol}, see discussion above), we arrived at the final parameterized values for $c$ and $d$, given by 0.04 K$^{-1}$ and -0.00037 K$^{-2}$ respectively (for details on the strategy for parameterization, refer to the SI). As discussed in the Methods section, these values are kept unchanged for all results shown in the rest of the manuscript.

%
\begin{figure}[!htb]
\includegraphics[width=0.49\textwidth]{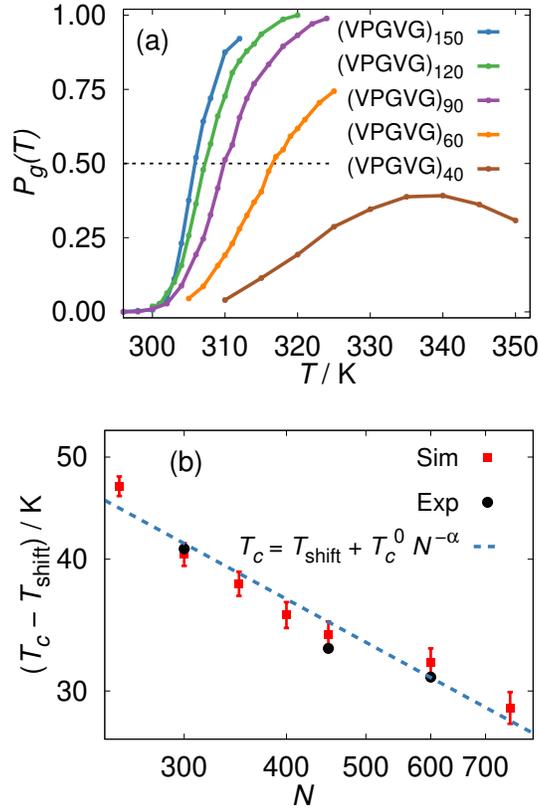}
\caption{{\footnotesize (a) Fraction of collapsed globule-like conformations ($P_g (T)$, eq~\ref{eqn:Tc}) as a function of temperature for (VPGVG)$_n$ sequences of varying pentameric repeat lengths $n$. The dotted line shows $P_g(T)=0.5$. (b) Estimates for transition temperatures ($T_c$) for (VPGVG)$_n$ sequences compared with experimental results by Meyer and Chilkoti~\cite{Meyer2004Biomacromol}. The experimental numbers correspond to lowest reported concentration of 1 $\mu$M. Log-log representation with a shift in $T_c$ is used to highlight that $T_c$ and sequence length $N$ (= $5n$) follow a power-law relation $T_c = T_{\mathrm{shift}} + T_c^0 N^{-\alpha}$ (see text), with $T_{\mathrm{shift}} \approx 276.1$ K, $T_c^0 \approx 461.5$ K and $\alpha \approx 0.42$ obtained from fit.}}
\label{fgr:LCST_ELPV}
\end{figure}
Following the parameterization of the temperature-dependence of our model using (VPGVG)$_{60}$ and (VPGVG)$_{90}$, we applied the model to (VPGVG)$_n$ sequences with pentameric repeats ($n$) in the broad range of 20-150. As is to be expected, the LCST transition occurs at progressively lower temperatures for larger $n$. This is shown in Figure~\ref{fgr:LCST_ELPV}(a) through the fraction of globular conformations in the ensembles ($P_g$, eq~\ref{eqn:Tc}) as a function of $T$. In Figure~\ref{fgr:LCST_ELPV}(a), $P_g(T_c)=0.5$ defines the $T_c$ for a given sequence length. As can be seen from Figure~\ref{fgr:LCST_ELPV}(b), $T_c\sim$ 308 K from simulations for (VPGVG)$_{120}$ is in excellent agreement with the experimental value of $\sim$ 307 K at 1 $\mu$M concentration obtained by Meyer and Chilkoti~\cite{Meyer2004Biomacromol}. For less than 50 pentameric repeats we do not observe LCST transitions in single chains, as shown for (VPGVG)$_{40}$ in Figure~\ref{fgr:LCST_ELPV}(a) where $P_g(T)$ approaches but never attains the value of 0.5. This is qualitatively consistent with reports from experimental literature, where at least $n$ = 40 pentameric repeats for (VPGVG)$_n$ have been required to observe LCST in dilute ELP solutions~\cite{Tatsubo2018,Kaibara1996,Maeda2013}. As such, we conclude that our model is able to accurately capture the LCST transition temperatures of (VPGVG)$_n$ sequences at low concentrations. 

In experiments, $T_c$ for ELPs have been observed to decrease with increasing ELP concentration following a logarithmic relationship~\cite{MacEwan2017,Meyer2004Biomacromol}. Experiments also reveal a power-law dependency of $T_c$ on chain length $N$ for thermoresponsive PNIPAM~\cite{PowerLawPNIPAM}. The latter dependence was also envisaged using all-atom simulations for ELPs~\cite{Zhao2016}. The most general form of a power-law dependence of $T_c$ on $N$ can be expressed as $T_c = T_{\mathrm{shift}} + T_c^0 N^{-\alpha}$, where $T_{\mathrm{shift}}$ is a reference correction. A fit of the functional form with $T_c$ values obtained from our simulations is shown in Figure~\ref{fgr:LCST_ELPV}(b). The values of $T_{\mathrm{shift}}$ and $T_c^0$ obtained from the fit are 276.1 K, and 461.5 K respectively. The exponent $\alpha$ is obtained to be $\approx$ 0.42, which is close to 0.44 reported for PNIPAM~\cite{PowerLawPNIPAM}, and not far from 0.5 predicted by theory~\cite{florybook,rubinsteinbook}.
\begin{table*}[!htb]
  \begin{center}
  \begin{tabular}{@{}ccccc@{}}
    \hline
    ELP   & $R_h^{exp}$ (293 K) & $R_h$ (293 K) & $R_h$ (340 K) & $R_{h,\mathrm{SOP-IDP}}$ (293 K)\\
    \hline
    (VPGVG)$_{150}$ & -- & 7.02 (0.32) & 3.46 (0.05) & 6.19 (0.29) \\
    (VPGVG)$_{120}$ & 6.0 & 6.12 (0.28) & 3.25 (0.05) & 5.52 (0.23)\\
    (VPGVG)$_{90}$ & -- & 5.22 (0.24) & 3.00 (0.06) & 4.83 (0.19) \\
    (VPGVG)$_{60}$ & 4.2 & 4.24 (0.19) & 2.69 (0.07) & 3.94 (0.15)\\
    (VPGVG)$_{40}$ & 3.7 & 3.49 (0.16) & 2.43 (0.09) & 3.21 (0.13)\\
    (VPGVG)$_{30}$ & 3.4 & 3.10 (0.14) & 2.28 (0.10) & 2.86 (0.12)\\
    (VPGVG)$_{20}$ & 2.7 & 2.56 (0.11) & 2.04 (0.09) & 2.37 (0.10)\\
    \hline
  \end{tabular}
  \end{center}
  \caption{{\footnotesize \textit{ELP chain dimensions:} Comparison of $R_h$ from simulations with reports by Schmidt \textit{et al.} using DLS experiments ($R_h^{exp}$)~\cite{Fluegel2010}. All numbers are in units of nm. The numbers in parentheses represent standard errors. The simulated $R_h$ values at 340 K are estimates for the dimensions of the most compact states, since $P_g(T)$ for the (VPGVG)$_n$ sequence shows maxima around 340 K (Figure~\ref{fgr:LCST_ELPV}(a)). The $R_{h,\mathrm{SOP-IDP}}$ column represents the $R_h$ values as obtained using the original, temperature-independent SOP-IDP model~\cite{SOPidp}.}}
  \label{tbl:Rh}
\end{table*}
\subsection{Structural properties of (VPGVG)$_n$: dimensions and disorder}
In the discussions above involving the parameterization of our model, attention was paid only to the distribution of $R_g$ values, without any regard to the mean dimensions of chains at any given temperature. In Table~\ref{tbl:Rh} we compare simulated hydrodynamic radii (eq~\ref{eqn:Rh}) of (VPGVG)$_n$ sequences at $T$ = 293 K with reported DLS measurements by Schmidt \textit{et al.} at the same temperature. For completeness and additional comparison, the results obtained using the original temperature-independent SOP-IDP model at the same simulation temperature are also shown in Table~\ref{tbl:Rh}. In Table S5 in the SI, we also report corresponding $R_h^{\mathrm{Nyg}}$ values calculated using the Nygaard relation~\cite{NygaardRh} and $R_g$ calculated from simulations (eq~\ref{eqn:Rg}).
First, given that the SOP-IDP model has been shown to capture the chain dimensions under ambient conditions for diverse polypeptide sequences~\cite{SOPidp,SOPidpABeta}, it is not surprising that the temperature-independent model results are already in fair agreement with experiments, with deviations in $R_h$ from the experimental values $\sim$(8-15)\%. It is gratifying to observe, that the temperature-dependent parameterization of the model results in improved agreement with experimental values, with maximum deviation now limited to $<$9\%. We reiterate that the improved accuracy in the reproduction of the chain dimensions is an emergent property from our model, and the temperature-dependent SOP-IDP model captures both \textquoteleft coil-like\textquoteright~ensemble dimensions, and the \textquoteleft coil-to-globule\textquoteright~transition temperatures of (VPGVG)$_n$ ELPs.

%
\begin{figure}[!htb]
\includegraphics[width=0.50\textwidth]{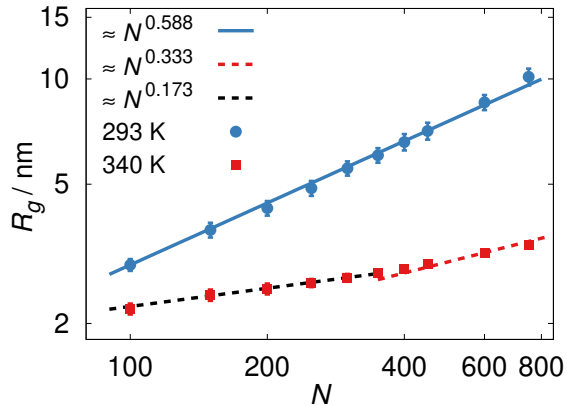}
\caption{{\footnotesize Scaling behavior of $R_g$~\cite{doiBook,Dobay03}, $R_g \sim R_g^0 N^{\nu}$, for (VPGVG)$_n$ sequences ($N = 5n$ is the sequence length),-- (blue) in the low temperature coil-like and (red) in the high temperature globule-like phases. The solid blue line reflects the good-solvent scaling at 293 K, with $\nu = 0.588$ and $R_g^0 \approx 0.2$ nm. The theoretically expected bad-solvent scaling is observed at 340 K for large sequence lengths (broken red line, $\nu = 0.333$ and $R_g^0 \approx 0.38$ nm). Smaller sequences at 340 K show a surprising weak scaling, shown by the broken black line, that fits best to $\nu \approx 0.173$, but with an unrealistic $R_g^0 \approx 1.0$ nm.}}
\label{fgr:RgScaling}
\end{figure}
Following the physics of polymers in good solvents, it is expected that the mean $R_g$ for (VPGVG)$_n$ at $T=293$ K should scale with with sequence length $N$ following Flory's scaling law $R_g = R_0^cN^{0.588}$~\cite{doiBook,Dobay03}. As shown in Figure~\ref{fgr:RgScaling}, this is indeed observed over the entire range in simulated sequence length from $N=100$ to $N=750$, with a prefactor $R_0^c \approx 0.2$ nm. Similarly, the $R_g$ values for the high $T$ globular states should scale as $R_g = R_0^gN^{1/3}$, if fully collapsed~\cite{doiBook}. From our simulations, we observed that the (VPGVG)$_n$ chains adopt most compact conformations at $T \sim$ (335 - 340) K, above which a weak expansion of the chains is observed. This observation is highly consistent with the \textit{unfolding regime} of $T \geq$ 333 K envisaged by Marx \textit{et al.}~\cite{Marx04}. As such, we consider the $R_g$ at 340 K as the \textit{globule} dimensions of single ELP chains ($R_h$ at 340 K are reported in Table~\ref{tbl:Rh}). Interestingly, the \textit{bad-solvent} scaling of $R_g \sim N^{1/3}$ only holds for $N\geq400$, with $R_0^g \approx 0.38$ nm, as shown also in Figure~\ref{fgr:RgScaling}. For smaller sequence lengths, $N < 350$, we clearly observe a weaker scaling. This is manifested through a systematic deviation towards larger $R_g$ than expected for the fully collapsed globular state of the polymers. 

%
\begin{figure}[!htb]
\includegraphics[width=0.5\textwidth]{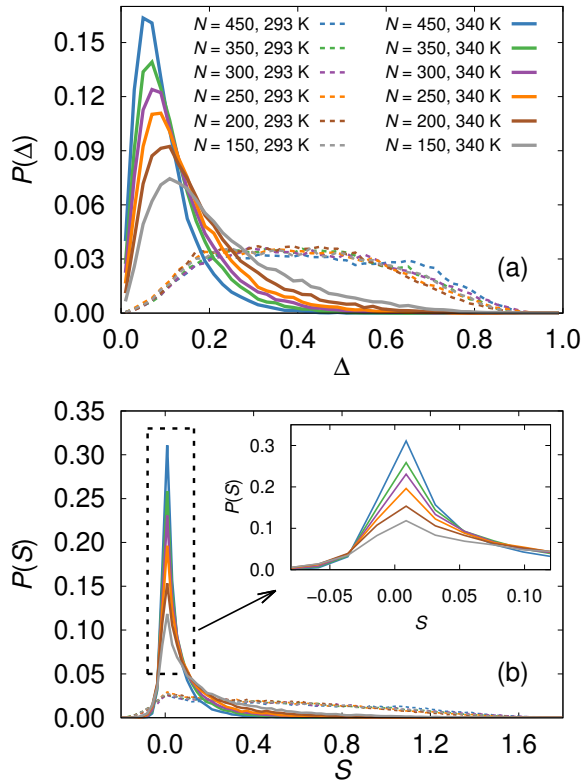}
 \caption{{\footnotesize Probability distributions of shape parameters $\Delta$ (a) and $S$ (b) for sequence lengths $150 \leq N \leq 450$ at 293 K (dashed) and 340 K (solid). The inset plot in (b) shows a zoomed-in view of the peaks around $S \sim 0$ at 340 K. }}
\label{fgr:PdeltaPs}
\end{figure}
Physical insights into the origins of this apparent weaker scaling can be gained through a statistical analysis of the shapes of the polymer conformations populating the ensembles of interest. Using the temperature-independent SOP-IDP model, and polymer shape parameters $\Delta$ and $S$ (eqs~\ref{eqn:shapeD},~\ref{eqn:shapeS})~\cite{ShapeRNA,Dima04JPCB}, it was shown that the conformational ensembles of disordered polypeptide sequences under good solvent conditions, specifically IDPs, can generically be described as prolate ellipsoidal, and with characteristically broad distributions in both $\Delta$ and $S$~\cite{SOPidp}. Predominantly prolate ellipsoidal conformations have also been reported in well solvated poly-ampholytic IDPs with low charge segregation along the sequence~\cite{DasProlate,DasProlate2}. Consistent with the observations, the distributions of shape parameters $\Delta$ and $S$ for (VPGVG)$_n$ sequences at $T$ = 293 K, shown with dashed lines in Figures~\ref{fgr:PdeltaPs}(a) and~\ref{fgr:PdeltaPs}(b), are observed to be broad, and highly skewed towards elongated prolate ellipsoidal conformations. For ready reference, we recall that $\Delta$ = 0 corresponds to a perfect sphere, while $\Delta$ = 1 is a rigid cylinder, and $S >$ 0 delineates prolate nature of an ellipsoid.

The overlapping nature of the distributions at $T$ = 293 K for all sequence lengths highlights the unchanged sequence composition with $N$. While the conformational ensembles are substantially more spherical at $T$ = 340 K for all $N$, a visual inspection of the $\Delta$ and $S$ distributions corresponding to different $N$ clearly show that the conformational ensembles become increasingly spherical as $N$ is increased from 150 to 450, with increasingly larger fractions of the conformations populating peaks corresponding to $\Delta<$ 0.2 and $S\sim$ 0. The $\Delta<$ 0.2 values indicate that the conformations are more spherical than the collapsed hairpin-like conformations observed in poly-ampholytic IDPs which are characterized by $\Delta \sim 0.2$~\cite{DasProlate}. 

%
Shape fluctuation is widely regarded as a key signature of conformational heterogeneity in disordered polypeptides, and is attributed to non-homogeneous distributions of amino acids along the peptide sequences~\cite{ConfHet1,ConfHet2,ConfHet3,Himadri,SOPidp}. Recently, the decoupling of size and shape fluctuations has been used to reconcile discrepancies in small-angle X-ray scattering and single-molecule F\"{o}rster resonance energy transfer measurements~\cite{ConfHet2}. The results in Figure~\ref{fgr:PdeltaPs} show that varying propensity for aspherical conformations can appear also through changes in sequence length. The block-wise repetitive sequences are identical in composition apart from the sequence length, and the environmental conditions for all simulations at 340 K are also identical. A key characteristic of the energy function at 340 K, in contrast to 293 K where shape distributions are invariant, is that it strongly promotes disproportionate interactions among monomers at short length scales. This likely results in locally non-isotropic packing of monomers. With increasing chain length, entropic constraints are relaxed, and the number of allowed pairwise contacts increases substantially. As a result, the ensembles sample greater numbers of near spherical conformations. Such considerations can be important in the interpretation of experimental data, especially in the quantification of aspherical deformations.

We conclude our results on the temperature-driven conformational changes in (VPGVG)$_n$ by highlighting the disordered nature of the ensembles,- both at low and high temperatures. While Figures~\ref{fgr:ELPV_param} and~\ref{fgr:RgScaling} already ascertain the good-solvent like polymeric nature and hence disorder at low $T$, the same is not trivially inferred at high $T$, especially for the longer chains showing LCST. Indeed, the $\sim N^{1/3}$ scaling of $R_g$ (Figure~\ref{fgr:RgScaling}, $N\geq400$) and relatively sharply peaked distribution of $R_g$ at 340 K (shown in Figure S5 in SI) can misleadingly be interpret as signatures of \textit{ordered} conformations, as predicted originally for ELPs. 
%
\begin{figure}[!htb]
\includegraphics[width=0.5\textwidth]{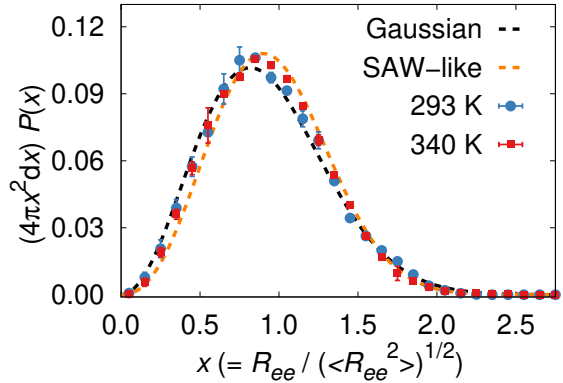}
 \caption{{\footnotesize Probability distributions of the scaled end-to-end distance ($R_{ee}$) for (VPGVG)$_{90}$ at $T$ = 293 K (blue circles) and $T$ = 340 K (red squares). The dashed curves in the background show the theoretical distributions for a Gaussian chain (black) and for a self-avoiding walk (orange), respectively~\cite{doiBook}. In our analyses, $R_{ee}$ for a polymer conformation is defined as the spatial distance between the terminal backbone beads in the chain.}}
\label{fgr:ReeDist}
\end{figure}
To provide a more stringent test for disorder in compact globular states (see also section titled \textit{Disordered ensemble and uncorrelated conformations} in the SI), we show in Figure~\ref{fgr:ReeDist} the scaled end-to-end distance ($R_{ee}$) distributions at both 293 K and 340 K for (VPGVG)$_{90}$. The distributions barely show any discernable changes, and agree reasonably well with distributions expected from theoretical polymer models. This reflects the existence of a dynamically evolving and highly mobile polymeric ensemble even in the most compact phase at $T$ = 340 K, above LCST. It is important to note, however, that this does not negate the possibility for transient local structures over short length scales.

\subsection{Relative hydrophobicities of amino acids}
%
%
\begin{figure}[!hb]
\vspace{-0.25in}
\includegraphics[width=0.50\textwidth]{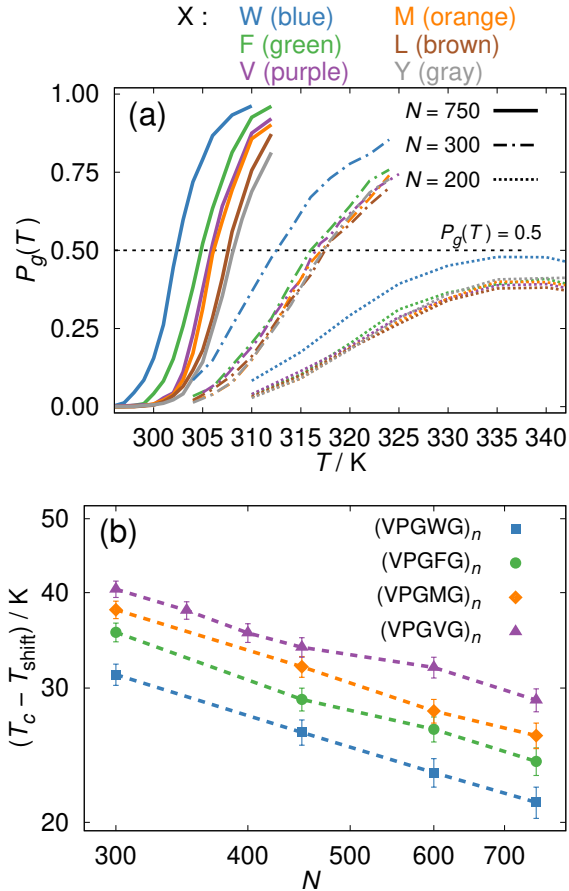}
\caption{{\footnotesize (a) Fraction of collapsed globule-like conformations ($P_g (T)$) as a function of $T$ for (VPG\textbf{X}G)$_n$ sequences, compared for sequence length $N$ (= $5n$) and guest residue identity (X = \{W, F, V, M, L, Y\}). The LCST transition can be observed to systematically sharpen for longer chains for all X. The residues W, F, V and M follow experimentally observed trends in relative hydrophobicities in the context of $T_c$ in ELPs, W>F>V$\sim$M~\cite{Urry1992Biopolymers}. Residues L and Y are less hydrophobic than expected. (b) Estimates for $T_c$ as a function of $N$. As with Figure~\ref{fgr:LCST_ELPV}(b), log-log representation is used to highlight the general power-law relation $T_c = T_{\mathrm{shift}} + T_c^0 N^{-\alpha}$. Using the $\alpha = 0.42$ reported in Figure~\ref{fgr:LCST_ELPV}(b), the fitted ($T_{\mathrm{shift}}$, $T_c^0$) values for (VPGWG)$_n$, (VPGFG)$_n$, (VPGMG)$_n$ and (VPGVG)$_n$ are (280.8 K, 347.9 K), (280.0 K, 392.3 K), (279.0 K, 422.9 K) and (276.1 K, 461.5 K) respectively. The fitted lines are not shown for clarity.}}
\label{fgr:LCST_ELPX}
\end{figure}
Following the initial works of Urry \textit{et al.}, and numerous experimental studies in subsequent years, it is understood that the chemical identity of the guest residue X in sequences (VPGXG)$_n$ influences the $T_c$ of the ELPs. More hydrophobic, especially aromatic-hydrophobic, residues lead to LCST transition at lower $T_c$, and \textit{vice-versa}~\cite{Urry1985Biopolymers,Urry1991JACS,Urry1992Biopolymers}. In the context of $T_c$ in ELPs (amino acid hydrophobicity scales, or rankings are context dependent and varied~\cite{HydSclTanford,HydSclKD,HydSclEis,HydSclZeb,HydSclBM,HydSclContAng}) the hydrophobic residues are ranked in descending order of hydrophobicities as Trp (W) > Tyr (Y) > Phe (F) > Leu (L) $\sim$ Ile (I) > Met (M) $\sim$ Val (V) > Ala (A)~\cite{Urry1992Biopolymers}. In the list, we have ignored the titratable residue histidine (H) for which the $T_c$ depends critically on the protonation state, and also proline (P), by definition in ELPs. The LCST transitions observed for the corresponding (VPGXG)$_n$ sequences are shown in Figures~\ref{fgr:LCST_ELPX}(a,b) for a diverse choice of X, and small ($n$ = 200) to long ($n$ = 750) sequence lengths. 

As shown in Figure~\ref{fgr:LCST_ELPX}(a), trends in the relative hydrophobicities of residues W, F, V, and M are faithfully reproduced by our model. The relative hydrophobicities are prominently observable for longer sequence lengths ($N$ = 750). For smaller sequence lengths the differences become increasingly less distinct -- with the exception of W, which is clearly observed as the most hydrophobic amino acid in our model. The lowest $T_c$ for guest residue W is immediately followed by a second aromatic-hydrophobic residue F, capturing the generally greater hydrophobicities of aromatic-hydrophobic residues compared to the aliphatic ones. The exception to this rule is tyrosine (Y), which is observed to be the least hydrophobic among the discussed hydrophobic amino acids except for alanine. Early studies report a unique characteristic of aromatic hydrophobic amino acids~\cite{Zehfus}, whereby depending upon methodology of characterization, they can be classified both as strongly hydrophobic~\cite{HydSclTanford} and mostly hydrophilic~\cite{HydSouthgate}. In solvent accessibility measurements of amino acids in folded proteins, Y residues show a low propensity for full burial unlike W and F~\cite{Zehfus}. The deviant behavior of Y in our model is attributed to this partial polar nature of Y, and consequent weaker interactions with hydrophobic residues in the knowledge-based Betancourt-Thirumalai potential used. This represents a limitation of the model in its current formalism. The complexity of the challenge dictates that the remedy should be found through incorporation of further specific interactions to the model, instead of a simple refinement of parameters. We refer the reader to a brief discussion in the Appendix. Following F are the hydrophobicities of V and M, which are virtually indistinguishable in our model. This is in reasonable agreement also with their hydrophobicity rank~\cite{Urry1992Biopolymers}. Deviating from the experimental trend, residues L and I have a higher $T_c$ than V and M. These deviations are discussed in the SI as part of context specific alterations of the model.

The general power law dependence of $T_c$ on sequence length $N$, discussed previously for (VPGVG)$_n$ is observed to hold for generic (VPGXG)$_n$ sequences, as can be seen from the approximately linear nature of the $\ln(T_c - T_{\mathrm{shift}})$ \textit{vs.} $\ln N$ plots shown in Figure~\ref{fgr:LCST_ELPX}(b). One can envisage that power law dependence would be observed for all polymers demonstrating LCST. From a computational perspective, it endows a degree of predictability to $T_c$ for a polymer model, which can be highly advantageous in efficient model building and parameterization.

%
%
\begin{figure*}[htb]
\includegraphics[width=\textwidth]{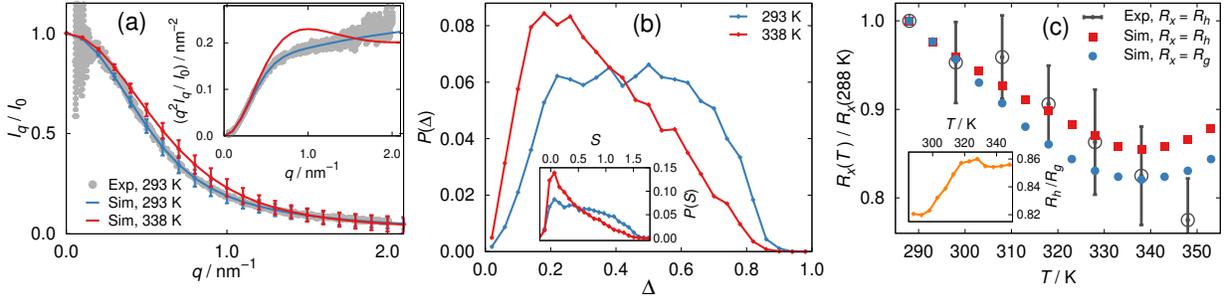}
\caption{{\footnotesize (a) Normalized scattering intensity (SAXS) profiles for the IDP p53-IDR. The grey profile is extracted from published experimental data at $T$ = 293 K by Fersht \textit{et al}~\cite{p53saxs}. The simulated profile at 293 K (blue) is in agreement with the experimental profile, and the simulated profile at 338 K (red) deviates strongly. Inset plot of (a) shows the same profiles in the Kratky representation. (b) Distributions of shape parameters $\Delta$ and $S$ (inset) from simulated conformations, at 293 K (blue) and 338 K (red) highlight the globular nature of the conformational ensemble of p53-IDR at 338 K, compared to 293 K. (c) The temperature-dependences of $R_h$ (red, square) and $R_g$ (blue, circle) for p53-IDR. Both quantities are plotted as the ratio of their corresponding values at $T$ = 288 K. Experimental data, extracted from published reference~\cite{p53DLS}, is shown in black, using mean values and error-bars. The inset panel of (c) shows the ratio $R_h/R_g$ at different simulated temperatures. Error bars in (c) were estimated using the propagation rule $\delta \left(\frac{x}{y}\right)/\left| \frac{x}{y} \right|\,=\,\frac{\delta \left(x\right)}{|x|} + \frac{\delta \left(y\right)}{|y|}$, with error margins in experimental $R_h(T)$ digitally extracted from published plots in cited reference. The error margins for corresponding simulated results are comparable to the experimental ones (Table S6 in SI), and have not been shown for clarity of representation. All experimental data were extracted digitally using an online tool \textit{WebPlotDigitizer}~\cite{webplot}.}}
\label{fgr:p53-IDR}
\end{figure*}

\subsection{Temperature-dependent conformational changes in IDPs}
In the results discussed thus far, we have elaborated on the performance of our model for simulations of ELP sequences (VPGXG)$_n$ over a wide range of experimentally relevant temperatures. To now challenge the performance of our model a bit further, we subject our simulations to study two intrinsically disordered polypeptide (IDP) sequences of substantially greater compositional complexity than ELPs. 

As discussed in the introduction, a variety of other, possibly temperature-dependent interactions contribute to the temperature-driven structural changes in IDPs. We recapitulate that in our simulations, in addition to $\epsilon(T)$, we consider only the temperature-dependence of solvent mediated electrostatic interactions (see methods). Our focus is thus directed at conformational changes induced by higher temperatures. Specifically, we simulated two IDPs,-- namely the 93 residue (1-93) intrinsically disordered region of protein p53 (p53-IDR) and the 441 residue human Tau protein (hTau40) over the wide temperature range of 288 K $\leq T \leq$ 348 K. The FASTA sequences for p53-IDR and hTau40 are provided in the SI. 

The IDPs p53-IDR and hTau40 are diverse in their amino acid compositions, having both a sufficiently high fraction of charged (20\% and 29\% respectively) and hydrophobic (60\% and 38\% respectively) amino acids. IDP p53-IDR, in spite of its relatively short sequence length, assimilates in its sequence all the hydrophobic amino acids discussed in this report. It has been studied experimentally at $T$ = 293 K using SAXS~\cite{p53saxs}, and over temperatures ranging from 278 K to 353 K using DLS and SDS-PAGE electrophoresis~\cite{p53DLS}. SAXS experiments have been reported with hTau40, encompassing a temperature range of 288 K to 333 K~\cite{saxsTauTdep,saxsTauTdep2012,saxsTau}. Both IDPs were studied previously in the development of the temperature-independent SOP-IDP model~\cite{SOPidp}. While neither is known to undergo LCST transition, they exhibit subtle temperature driven compaction of their overall dimensions~\cite{saxsTauTdep,saxsTauTdep2012,p53DLS}.\\

\noindent
\textbf{p53-IDR:} Figure~\ref{fgr:p53-IDR}(a) compares simulated SAXS scattering intensity profiles (eq~\ref{eqn:Debye}) with experimental SAXS profile reported by Fersht \textit{et al.} at 293 K~\cite{p53saxs}. The simulated profile at $T$ = 293 K is in agreement with the experiment, both in the normalized scattering intensity ($I_q/I_0$) and the Kratky ($q^2I_q/I_0$) representation (inset plot, Figure~\ref{fgr:p53-IDR}(a)). Quantitative consistency over the full experimentally reported range in $q$ establishes that the simulated ensemble is a highly faithful representation of the conformational ensemble of p53-IDR at the low temperature condition of $T$ = 293 K. As can be expected, the $R_g$ at 293 K computed from our current simulation data using eq~\ref{eqn:Rg}, 2.92 $\pm$ 0.5 nm, is also in good agreement with the experimentally reported value of 2.87 nm (2.95 nm at 298 K was obtained with the temperature-independent SOP-IDP model)~\cite{SAXSRevSvergun,SOPidp}. 

The simulated SAXS profile at $T$ = 338 K is also shown in Figure~\ref{fgr:p53-IDR}(a) using both scattering intensity and Kratky representations. Clear deviations from the corresponding curves at 293 K indicates a substantial dependence of the conformational ensemble of p53-IDR on temperature. Of special interest here is the appearance of the bell-shaped nature of the curve in the Kratky representation at 338 K, which implies a propensity towards more compact, globule-like conformations~\cite{SAXSguidePractical}. The polymer shape parameters $\Delta$ and $S$ distributions at the two temperatures are shown in Figure~\ref{fgr:p53-IDR}(b). Both distributions clearly highlight that the conformational ensemble shifts from extended ellipsoidal conformations at 293 K to comparatively globular conformations at 338 K. The simulated $R_g$ at 338 K for p53-IDR was observed to be 2.47 $\pm$ 0.5 nm, reduced by $\sim 15\%$ from its value at 293 K.

Hydrodynamic radius measurements for p53-IDR have been reported by Whitten \textit{et al.} over a wide range of temperatures~\cite{p53DLS}. The simulated ($R_g$, $R_h$) and experimental ($R_h$) size estimates are reported in Table S6 in the SI as a function of temperature. Table S6 also includes $R_h$ calculated from our simulations using the Nygaard relation~\cite{NygaardRh}. In spite of good agreement with SAXS described above, and the known ability of the SOP-IDP model to capture both $R_g$ and $R_h$ for other IDPs (\textit{e.g.} $\alpha$-Synuclein) to acceptable accuracy~\cite{SOPidp}, we observed that the reported $R_h$ values were systematically larger than the simulated values, even around 293 K. A closer inspection of reported experimental values reveals an $R_h/R_g$ ratio $>$ 1.1 around 293 K, which deviates strongly from theoretical polymer models (limiting values of 0.665 and 0.640 for ideal chain and polymer in a good solvent, respectively~\cite{TeraokaBook}) and observations from simulations of IDPs~\cite{SOPidp}. As shown in the inset of Figure~\ref{fgr:p53-IDR}(c), the simulated value of $R_h/R_g$ initially increases with $T$ from $\sim$ 0.82 and saturates around $\sim$ 0.86, in agreement with prior observations with disordered states of IDPs~\cite{SOPidp}. To compare our simulated results with experiments by Whitten \textit{et al.}, we thus focus on the compaction of the normalized size of p53-IDR with temperature. The respective size estimates (experimental and simulated) at $T$ = 288 K act as appropriate normalizations.

Figure~\ref{fgr:p53-IDR}(c) compares $R_h(T)/R_h(288K)$ from DLS measurements~\cite{p53DLS} and our simulations. For completeness, $R_g(T)/R_g(288K)$ data from simulations is also shown. The comparison shows good qualitative agreement in temperature induced compaction over a wide range of temperatures ranging from 293 K (agreement at 288 K holds trivially through construction) to $\sim$ 338 K, with simulated $R_h(T)/R_h(288K)$ data points always found within the experimental error margins in the said interval. Promising agreement is found in the smaller temperature range of 293 K $\leq T \leq$ 328 K. Beyond the high temperature of $T$ $\sim$ 340 K, the simulated results show an increase in size instead of compaction. In our simulations, this is attributable to the peaking of pairwise hydrophobic interactions $\epsilon(T)$ around $T \sim$ 340 K as discussed in previous sections. We refer the reader to the Appendix section, where we provide further discussion on this observed deviation. \\
\begin{table}[!htb]
  \begin{center}
  \begin{tabular}{@{}cccc@{}}
    \hline
    $T$ / K & $R_g^{exp}$ / nm & $R_g^{sim}$ / nm & \% deviation\\
    \hline
    293 & 6.96 (0.16) & 6.77 (0.81) & 2.73 \\
    298 & 7.00 (0.17) & 6.56 (0.80) & 6.29 \\
    303 & 6.53 (0.15) & 6.38 (0.75) & 2.30 \\
    308 & 6.40 (0.16) & 6.22 (0.76) & 2.81 \\
    313 & 6.34 (0.17) & 6.16 (0.69) & 2.84 \\
    318 & 5.93 (0.20) & 6.08 (0.79) & 2.53 \\
    323 & 6.06 (0.20) & 6.01 (0.78) & 0.83 \\
    328 & 5.90 (0.21) & 5.94 (0.72) & 0.68 \\
    333 & 5.67 (0.23) & 5.99 (0.82) & 5.64 \\
    338 & -- & 6.07 (0.80) & -- \\
    343 & -- & 6.20 (0.81) & -- \\
    \hline
  \end{tabular}
  \end{center}
  \caption{{\footnotesize Comparison of experimentally reported ($R_g^{exp}$)~\cite{saxsTauTdep,saxsTauTdep2012} and simulated ($R_g^{sim}$) radii of gyration for IDP hTau40 at different temperatures. Numbers in parentheses represent error estimates. The \% deviation in the final column is calculated as $\left( \left|R_g^{exp} - R_g^{sim}\right| / R_g^{exp}\right) \times 100\%$. Experimental values were obtained, as reported in cited reference, using 1000 representative hTau40 conformations generated through ensemble optimization method fits of the scattering intensity profiles at each temperature~\cite{saxsTauTdep}.}}
  \label{tbl:RgTau}
\end{table}

%
\begin{figure}[htb]
\vspace{-0.25in}
\includegraphics[width=0.50\textwidth]{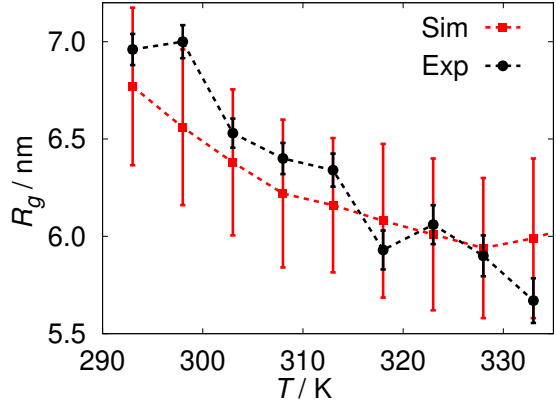}
\caption{{\footnotesize Visual comparison of experimental~\cite{saxsTauTdep,saxsTauTdep2012} (black circles) and simulated (red squared) $R_g$ data for IDP hTau40 listed in Table~\ref{tbl:RgTau}}.}
\label{fgr:RgTau}
\end{figure}
\begin{figure*}[!htb]
\includegraphics[width=\textwidth]{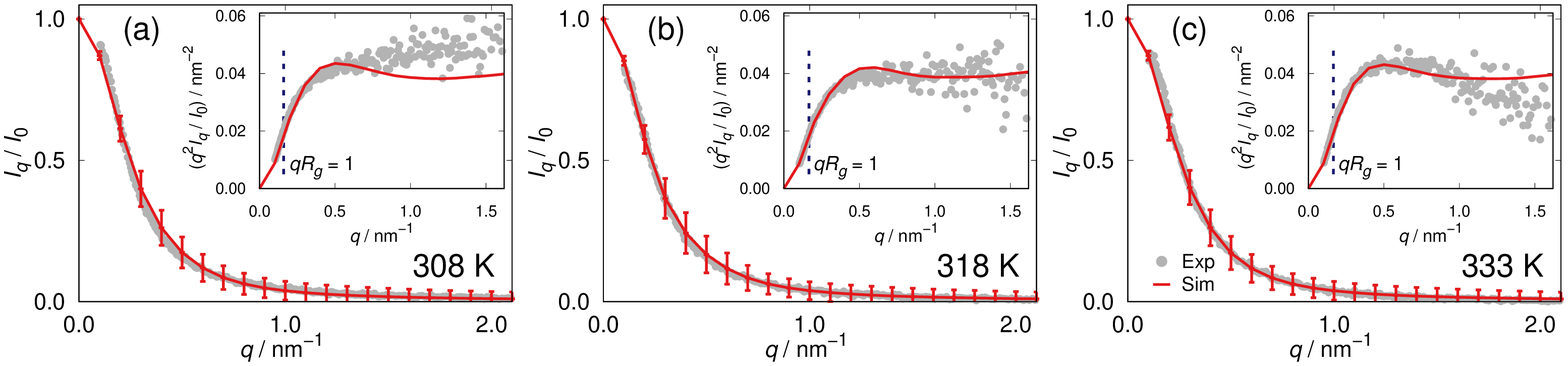}
\caption{{\footnotesize Normalized scattering intensity (SAXS) profiles for the IDP hTau40 at (a) 308 K, (b) 318 K and (c) 333 K. Simulated profiles are shown in red. Experimental data~\cite{saxsTauTdep,saxsTauTdep2012}, shown in grey, were received courtesy of \emph{A. Tenenbaum} and \emph{G. Ciasca}. Inset plots show the profiles in the Kratky representation. The broken vertical lines in the inset plots show the respective Guinier regimes ($qR_g \leq 1$) using $R_g=R_g^{sim}$ from Table~\ref{tbl:RgTau}.}}
\label{fgr:SAXSTau}
\end{figure*}
\noindent
\textbf{hTau40:} The 441 residue long hTau40 is among the longest IDP sequences whose single chain conformational properties have been studied till date in experiments~\cite{saxsTau,saxsTauTdep,SAXSRevSvergun,SOPidp}. It has been shown to undergo a temperature induced compaction in $R_g$ by $\sim$ 18\% between the temperatures of 293 K and 333 K using SAXS measurements by Tenenbaum \textit{et al}~\cite{saxsTauTdep,saxsTauTdep2012}. The $R_g$ from our simulations at different simulation temperatures are compared with the corresponding experimental values in Table~\ref{tbl:RgTau} and Figure~\ref{fgr:RgTau}. 
As the comparisons reflect, our simulation model not only captures a sufficiently faithful representation of the temperature-driven fractional compaction of hTau40 ($\sim$ 12\% from simulations \textit{vs.} $\sim$ 15\% from experiments between 293 K and 328 K), but also consistently reproduces the absolute experimental $R_g$ values to appreciable accuracy. For all temperatures apart from 298 K and 333 K, the deviations from experimental values is limited to $<$ 3\%. Approaching high $T$ values $\geq$ 333 K, the simulation results show thermal expansion as discussed previously in the manuscript, in contradiction to continued compaction reported by experiment at 333 K (Figure~\ref{fgr:RgTau}). 

SAXS scattering intensity profiles ($I_q$, eq~\ref{eqn:Debye}) enable direct comparison between experimental and simulated conformational ensembles over varying length-scales ($q^{-1}$). In Figure~\ref{fgr:SAXSTau} we compare simulated and experimental~\cite{saxsTauTdep,saxsTauTdep2012} $I_q$ profiles at three above-ambient temperatures. Consistent with the agreement observed in $R_g$, the simulated and experimental profiles are in good agreement beyond the Guinier regime ($qR_g \leq 1$). Indeed, over a wide range of temperature, strong deviations from the experimental profiles are observed only beyond $q \sim 5 R_g^{-1}$, as highlighted in the Kratky representation (inset plots). At $T \sim 318$ K the simulated ensemble appears to be a faithful representation of the experimental ensemble (Figure~\ref{fgr:SAXSTau}(b)). For a lower temperature of $T \sim 308$ K (Figure~\ref{fgr:SAXSTau} (a)), over-compaction is observed in the simulated ensemble. The scenario is reversed at $T \sim 333$ K, with the experimental ensemble appearing to be locally compact over the simulated ensemble. With its complex sequence composition and associated conformational heterogeneity, IDP hTau40 has historically been a challenging system for study, even at ambient temperatures~\cite{saxsTau,SOPidp}. While clearly imperfect, the results are promising, and detailed analysis of the temperature-dependent conformations of hTau40 will be part of future projects. 
\section{Concluding remarks}
We have presented an explicitly temperature-dependent coarse grained model for the simulations of disordered, predominantly hydrophobic polypeptide sequences. The model captures the structure and loci of the cloud point (here, collapse transition temperature) for a variety of experimentally probed ELP sequences, most notably for the \textit{canonical} ELPs (VPGVG)$_n$. We provided insights into the temperature-dependent size-scalings, in particular the shape behavior of globular conformations at high temperatures which indicate an anisotropic nature of the collapsed state for lower degrees of polymerization. The model is shown to capture also the high temperature compaction of more complex polypeptides, namely the IDPs p53-IDR and hTau40. A detailed study of other IDPs using the model, and the structure of their high temperature coacervates would be very interesting for future directions. Finally, we envisage that our model will be highly useful in future simulations and characterization of ELP networks and hydrogels across the collapse transition, which is very important for the development of highly responsive, functional soft materials~\cite{Andreas}.

\begin{acknowledgements}
UB would like to thank S\'{e}bastien Groh for insightful discussions. The authors are indebted to A. Tenenbaum and G. Ciasca for tabulated experimental scattering profiles of hTau40. This project has received funding from the European Research Council (ERC) under the European Union's Horizon 2020 research and innovation programme (grant agreement no. 646659). The authors acknowledge support by the state of Baden-W\"urttemberg through bwHPC and the German Research Foundation (DFG) through grant no INST 39/963-1 FUGG (bwForCluster NEMO). 
\end{acknowledgements}
\section{Appendix : Limitations of model and possible remedy}
In the article we have taken care to describe the known limitations of the model. In the following, we briefly outline some possible approaches for improvement, which also represent possible future directions of research. 

\emph{Deviations of relative hydrophobicities of amino acids from experiments:} The small deviations observed for aliphatic amino acids can be easily accounted for, as described in the SI. We believe that the low observed hydrophobicity of Y should not be corrected purely within the scope of the model. Aromatic hydrophobic residues can contribute to additional residue-specific interactions such as $\pi$-$\pi$ or cation-$\pi$ interactions, and hydrogen bonding. These interactions are stronger at lower temperatures, and are known to play important roles in UCST behavior of polypeptides~\cite{RevMittag2018BioChem}. In future work, we intend to extend the model to UCST transitions by including the said interactions, possibly in temperature-dependent manner. While specific results can not be predicted, presence of stronger interactions at low temperatures will drive the $T_c$ of Y containing sequences to lower temperatures. This effect should be accounted for before attempting to optimize of interaction parameters for Y. 

\emph{Increase in simulated size of IDPs at very high temperatures:} We have reported small expansions in simulated sizes of IDPs p53-IDR and hTau40 around and above $T \sim 333$ K, which contradict continued compactions reported in experiments. The maxima of pairwise $\epsilon(T)$ is strongly sensitive to parameters in eq~\ref{eqn:EpsT}. Thus, minor modifications to a parameter such as $d$ in eq~\ref{eqn:EpsT} is a possible direct route to rectify these deviations. However, the current model does not account for temperature-dependence of pairwise interactions among polar and charged (beyond screened electrostatics) residue types. The strength of these pairwise interactions can increase at very high temperatures, driven by the reduced stability of hydration shell structures around these residues. It is possible that the increased strength of such interactions also contributes to the continued compactions of the IDPs beyond $T \sim 333$ K. Existing literatures suggest that temperature-dependent modifications to such pair interactions in a simulation model should be introduced through changes in vdW radii, rather than $\epsilon(T)$~\cite{ElcockAA,Chudoba2017PEG}.
%
\bibliography{ELP}
%

\end{document}